\documentclass[]{aa}

\usepackage{amsmath,amssymb}
\usepackage{graphicx}
\usepackage{natbib}

\newcommand{\elie}{{\mathbf E}}

\newcommand{\ain}{a_{\rm in}}
\newcommand{\aout}{a_{\rm out}}
\newcommand{\Sigin}{\Sigma_{\rm in}}
\newcommand{\Sigout}{\Sigma_{\rm out}}

\newcommand{\elik}{{\mathbf K}}

\newcommand{\psiin}{\psi_{\rm in}}
\newcommand{\psiout}{\psi_{\rm out}}

\newcommand{\psiapp}{\psi_{\rm app.}}

\begin{document}

\title{A new equation for the mid-plane potential of power law discs}
\titlerunning{A new equation for the mid-plane potential of power law discs. II Exact solutions}

\subtitle{II. Exact solutions and approximate formulae}

\author{Jean-Marc Hur\'e\inst{1,2}, Franck Hersant\inst{2}, Cyril Carreau\thanks{Present address : La Maurellerie, 37290 Bossay-sur-Claise} \and Jean-Pierre Busset\inst{1}}
\authorrunning{Hur\'e et al.}

\offprints{Jean-Marc Hur\'e}

\institute{Universit\'e de Bordeaux, LAB, 351 cours de la Lib\'eration, Talence, F-33405, France
\and
CNRS/INSU, UMR 5804/LAB, 2 rue de l'Observatoire, BP 89, Floirac Cedex, F-33271, France\\
\email{jean-marc.hure@obs.u-bordeaux1.fr}\\
\email{franck.hersant@obs.u-bordeaux1.fr}\\
\email{cyril.carreau@wanadoo.fr}\\
\email{jean-pierre.busset@obs.u-bordeaux1.fr}
}

\date{Received ??? / Accepted ???}

\abstract
{}
{The first-order ordinary differential equation (ODE) that describes the mid-plane gravitational potential in flat finite size discs of surface density $\Sigma(R) \propto R^{s}$ (Hur\'e \& Hersant 2007) is solved exactly in terms of infinite series.}
{The formal solution of the ODE is derived and then converted into a series representation by expanding the elliptic integral of the first kind over its modulus before analytical integration.}
{Inside the disc, the gravitational potential consists of three terms: a power law of radius $R$ with index $1+s$, and two infinite series of the variables $R$ and $1/R$. The convergence of the series can be accelerated, enabling the construction of reliable approximations. At the lowest-order, the potential inside large astrophysical discs ($s \sim -1.5 \pm 1$) is described by a very simple formula 
whose accuracy (a few percent typically) is easily increased by considering successive orders through a recurrence. A basic algorithm is given.
}
{Applications concern all theoretical models and numerical simulations where the influence of disc gravity must be checked and/or reliably taken into account.}
{}


\maketitle

\section{Introduction}

Gaseous discs in which the main physical quantities (density, pressure, temperature, thickness, velocity) scale with cylindrical radius as power laws, i.e. ``power-law discs'', represent an important class of theoretical systems. These are used customary to model accretion in evolved binaries \citep{ss73,pringle81}, circumstellar matter \citep{dubrulle92,edgar07}, the environment of massive black holes inside active galactic nuclei \citep{cd90,hure98,semerak04} or even the stellar component of some galaxies \citep{evans94,zhao99}. In most applications however, power-law discs are truncated either to avoid diverging values at the disc centre (such as density, mass) or in attempting to reproduce the properties of observed discs of finite extension and mass. Although self-similarity is not compatible with the presence of edges, it is generally considered that power laws offer a good description of disc properties in some regions (far from the edges).  Note that the presence of sharp edges can be misleading when interpreting observational data \citep[e.g.][]{hughes08}. In general, the surface density $\Sigma$ in the outer parts of discs is a decreasing function of the cylindrical radius $R$. Depending on the models, hypotheses, and objects, we have, for instance, $\Sigma \propto R^{-3/5}$ in binaries  \citep{ss73}, $\Sigma \propto R^{\pm9/20}$ in active galactic nuclei \citep{cd90}, $\Sigma \propto R^{-1}$ for a Mestel disc \citep{mestel63}, or $\Sigma \propto R^{-3/2}$ in circumstellar discs \citep{pietu07}. In the context of stationary viscous $\alpha$-discs, a wide range of power-law exponents is allowed since the temperature $T$, $\Sigma$, and $R$ satisfy the condition \citep[e.g.][]{pringle81}:
$$\Sigma T R^{3/2} =cst,$$
while it is $\Sigma \propto R^{-1/2}$ in $\beta$-discs \citep{hrz01}.\\

The calculus of the gravitation potential of finite-size, power-law discs has received little attention yet. Several reasons can be put forward. Solving the Poisson equation or computing the integral of the potential is not a trivial procedure, especially in the presence of edges. It is generally believed that gravity due to low mass discs is unimportant compared with that of a central proto-star or black hole, and cannot be probed \citep[see however][]{baruteaumasset08}. Many studies employ the multi-pole expansion which is known to converge too slowly inside sources to be efficient for the numerical applications \citep[e.g.][]{clement74,stonenorman92}. \cite{hh07} demonstrated that the mid-plane potential of flat power-law discs obeys an inhomogeneous first-order Ordinary Differential Equation (ODE). In this second paper, we discuss the exact solutions of this ODE for the entire physical range (outside and inside the disc) in terms of infinite series. In particular, it is shown that the mid-plane potential is a combination of a power law for the radius $R$ and two series of the variables $R$ and $1/R$. Since these series converge rapidly inside large discs, it is possible to derive reliable approximations by truncating the series at low orders.\\

This paper is organised as follows. The ODE for the potential is briefly recalled in Sect. \ref{sec:unified} and its formal solution is derived in Sect. \ref{sec:Formal solution of the ODE}. In Sect. \ref{sec:pate}, we express the potential at the two disc edges and consider a few special cases. The inside and outside solutions in the form of series are presented in Sect. \ref{sec:Solution of the ODE}. In Sect. \ref{sec:Potential inside the disc}, we analyse the potential in the disc inside in detail, and in particular, the power-law contribution. Since all series involved converge rapidly, we are able to derive reliable approximations for the potential; this is done in Sect. \ref{sec:Approximate formulae}. We discuss in Sect. \ref{sec:Discs with no inner/outer edges} the case of discs with no inner and/or outer edge. The paper ends with a few concluding remarks.

\begin{figure}[h]
\includegraphics[width=9.0cm]{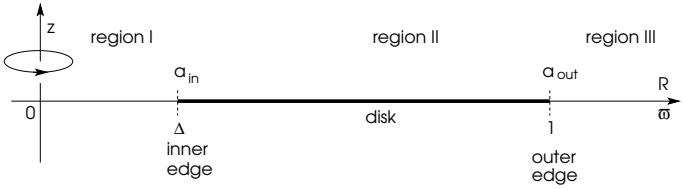}
\caption{Configuration for a finite-size flat disc.}
\label{fig:scheme.eps}
\end{figure}     

\section{The mid-plane potential in power-law discs from an Ordinary Differential Equation (ODE)}
\label{sec:unified}

Following \cite{hh07} (hereafter Paper I), the mid-plane potential $\psi$ due to a flat power-law disc satisfies the ODE:
\begin{equation}
\frac{d \psi}{d \varpi} - (1+s)\frac{\psi}{\varpi} = S(\varpi),
\label{eq:compactode}
\end{equation}
where  $\varpi=R/\aout$ is the cylindrical radius in units of the radius $\aout$ of the disc outer edge, $s$ is the power-law index of the surface density $\Sigma$, namely (generally, $s < 0$ in astrophysical discs):
\begin{equation}
\Sigma \propto R^s,
\end{equation}
and $S(\varpi)$ is the piecewise defined function. Depending on the position in the disc (see Fig. \ref{fig:scheme.eps}), we have:
\begin{equation}
\label{eq:spiecewise}
S(\varpi) = \frac{ 2 \psiout}{\pi \varpi^2}
\begin{cases}
\varpi \left[ \elik(\varpi) - \elik\left(\frac{\varpi}{\Delta}\right)\Delta^{1+s}  \right]\\ \quad {\rm for} \quad 0 \le \varpi \le \Delta \qquad \text{(region I)},\\\\
  \varpi \elik(\varpi) - \elik\left(\frac{\Delta}{\varpi}\right) \Delta^{s+2}\\ \quad {\rm for} \quad \Delta \le \varpi \le 1 \qquad \text{(region II)},\\\\
 \elik\left(\frac{1}{\varpi}\right) - \elik\left(\frac{\Delta}{\varpi}\right)\Delta^{s+2}\\ \quad {\rm for} \quad \varpi \ge 1 \qquad \text{(region III)},
\end{cases}
\end{equation}
where $\Delta=\ain/\aout < 1$ is the axis ratio, $\ain$ is the radius of the inner edge, $\psiout=2 \pi G \Sigout \aout$ is a positive constant, $\Sigout$ is the surface density at the disc outer edge, $\elik$ is the complete elliptic integral of the first kind:
\begin{equation}
\elik(x)=\int_0^{\pi/2}{\frac{d \phi}{\sqrt{1-x^2 \sin^2 \phi}}}, \quad 0 \le x \le 1,
\end{equation}
and $G$ is the gravitation constant. The above ODE can in principle be solved in the entire radial domain since boundary conditions (both $\psi$ and the associated acceleration $-d_\varpi \psi$) are known precisely at $\varpi=0$ and at $\varpi=\infty$ for power law distributions. Although $S$ is singular at the two disc edges (i.e. for $\varpi \in \{\Delta,1\}$), Eq. (\ref{eq:compactode}) is far more tractable in computing $\psi$ than the integral form \citep[e.g.][]{durand64}:
\begin{equation}
\psi(\varpi) = -4 G  \Sigout \aout^{-s} \int_{\ain}^{\aout}{\frac{a^{s+1}}{a+R} \elik\left(\frac{2\sqrt{aR}}{a+R}\right)da},
\label{eq:psiintegralform}
\end{equation}
whose integrand is logarithmically singular {\it everywhere inside the disc}, or than the Poisson equation, which involves vertical gradients.
                                                               
\section{Formal solution of the ODE}
\label{sec:Formal solution of the ODE}

A formal solution of Eq. (\ref{eq:compactode}) is found by setting \citep[e.g.][]{rybicki79}:
\begin{equation}
\bar{\psi}(\varpi) = \varpi^{-(1+s)} \psi(\varpi),
\label{eq:phi}
\end{equation}
and
\begin{equation}
\bar{S}(\varpi) = \varpi^{-(1+s)} S(\varpi).
\label{eq:sigma}
\end{equation}
The exact derivative of $\bar{\psi}$ is:
\begin{equation}
\frac{d \bar{\psi}}{d \varpi} = \varpi^{-(1+s)}\left[\frac{d \psi}{d \varpi} - (1+s)\frac{\psi}{\varpi}\right]
\end{equation}
where we recognise, inside brackets, the function $S$. Therefore, we have:
\begin{equation}
\frac{d \bar{\psi}}{d \varpi} = \bar{S}(\varpi),
\end{equation}
whose formal solution for $\bar{\psi}$ is of the form:
\begin{equation}
\bar{\psi}(\varpi) = \bar{\psi}(\varpi_0) + \int_{\varpi_0}^\varpi{\bar{S}(\varpi')d\varpi'},
\end{equation}
Back-substituting $\bar{\psi}$ and $\bar{S}$ from Eqs.(\ref{eq:phi}) and (\ref{eq:sigma}), we find the general expression for the mid-plane potential:
\begin{equation}
\psi(\varpi) = \varpi^{1+s} \left[ \frac{\psi(\varpi_0)}{\varpi_0^{1+s}} + \int_{\varpi_0}^\varpi{\frac{S(\varpi')}{{\varpi'}^{1+s}}d\varpi'} \right].
\label{eq:formals}
\end{equation}
This solution is fully determined in the entire spatial domain or part of it as soon as the potential is known at a given normalised radius $\varpi_0$. We observe that, for $s \ne -1$, $\psi(\varpi)$ is the mixture of a power law of the radius (the first term in the right-hand-side) with exponent \citep[e.g.][]{bkogan75,evansread98}:
\begin{equation}
p = 1+s,
\label{eq:p}
\end{equation}
and a complicated function of the radius $R$ (the definite integral).

In the following, we shall analyse Eq. (\ref{eq:formals}) analytically in terms of infinite series by considering in Eq. (\ref{eq:spiecewise}) the expansion of $\elik$ over its modulus \citep[e.g.][]{gradryz65}:
\begin{equation}
\left\{
\begin{aligned}
&\elik(x)= \frac{\pi}{2}  \sum_{n=0}^{\infty}{\gamma_n x^{2n}} \quad \text{with} \quad 0 \le x \le 1,\\ \\
&\gamma_0 =1,\\ \\
&\gamma_n = \gamma_{n-1} \left(\frac{2n-1}{2n}\right)^2, \qquad n \ge 1.
\label{eq:kseries}
\end{aligned}
\right.
\end{equation}

\section{Potential at the disc edges}
\label{sec:pate}

\subsection{Inner edge}

To determine $\psi(\varpi)$ from Eq. (\ref{eq:formals}) for $\varpi \in [0,\infty[$, it is sufficient to calculate the potential at the two disc edges\footnote{In contrast with the gravitational acceleration (the gradient of $\psi$), the potential is generally finite at the edges. \label{note:ktrans}}, that is $\psi(\Delta)$ and $\psi(1)$. For this purpose, we use Eq. (\ref{eq:psiintegralform}) and define $v = R/a \le 1$. After some algebra\footnote{In particular, we use the transformation:
\begin{equation}
\elik\left(\frac{2\sqrt{x}}{1+x}\right) = (1+x) \elik(x), \quad 0 \le x \le 1.
\label{eq:ktrans}
\end{equation}} we find:
\begin{equation}
\psi(\Delta) = 4G \Sigout \Delta^{1+s} \aout \int_1^{\Delta}{\frac{\elik(v)}{v^{2+s}}dv}.
\label{eq:psiin}
\end{equation}

From Eq. (\ref{eq:kseries}), the potential at the inner edge is:
\begin{equation}
\psi(\Delta) = -\psiout \Delta^{1+s} \sum_{n=0}^\infty{I_n \left(1-\Delta^{2n-s-1}\right)},
\label{eq:psiin2}
\end{equation}
where
\begin{equation}
I_n = \frac{\gamma_n}{2n-s-1},
\end{equation}
and assuming $\Delta \ne 0$ (see below). As $\gamma_n \sim 1/n$ for large $n$ (see Fig. \ref{fig:spectredn.eps}), $\psi(\Delta)$ consists of terms that vary asymptotically like $\sim 1/n^2$. Since $\Delta < 1$, $\psi(\Delta)$ is a converging series.

\begin{figure}
\includegraphics[width=9.0cm]{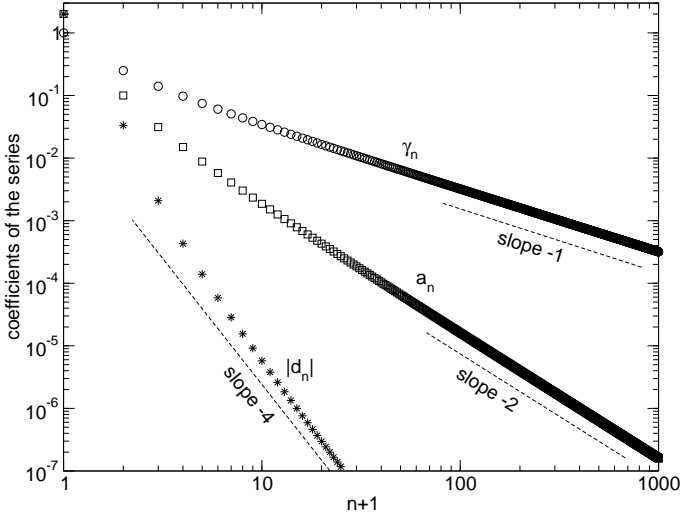}
\caption{Coefficient $\gamma_n$ versus $n$. Terms $a_n$ and $|d_n|$ versus $n$ for $s=-1.5$.}
\label{fig:spectredn.eps}
\end{figure}

\subsection{Outer edge}

At the outer edge, the potential is calculated in a similar manner, but using the variable $u = a/R \le 1$. For $\varpi=1$, Eq. (\ref{eq:psiintegralform}) writes$^{\ref{note:ktrans}}$:
\begin{equation}
\psi(1) = - 4G \Sigout \aout \int_\Delta^1{\elik(u)u^{1+s}du}.
\label{eq:psiout}
\end{equation}

Replacing $\elik(u)$ by its series representation yields:
\begin{equation}
\psi(1) = - \psiout \sum_{n=0}^\infty{J_n \left(1 - \Delta^{2n+s+2}\right)},
\label{eq:psiout2}
\end{equation}
where
\begin{equation}
J_n = \frac{\gamma_n}{2n+s+2},
\end{equation}
and by assuming $\Delta \ne 0$ (see below). As for $\psi(\Delta)$ and for the same reasons, $\psi(1)$ is also a converging series.

\subsection{Special values of $s$}
\label{subsec:specs}

If the power law exponent $s$ is such that:
\begin{equation}
2n-s-1=0
\label{eq:ndelta}
\end{equation}
 at a certain rank $n \equiv n_\Delta$, then $\psi(\Delta)$ must be, in practice, written in a slightly different form. This happens for $s \in {\cal E}_\Delta$ with:
\begin{equation}
{\cal E}_\Delta = \{-1,+1,+3,+5,\dots \}.
\end{equation}
Since
\begin{equation}
\label{fn:div}
\lim_{q \rightarrow 0 }\frac{1 - x^q}{q} = -\ln x,
\end{equation}
for $x >0$ and any $q$, we have:
\begin{equation}
\lim_{n  \rightarrow n_\Delta }I_n(1 - \Delta^{2n-s-1}) = -\gamma_{n_\Delta}\ln \Delta.
\end{equation}

Then, if $s \in {\cal E}_\Delta$, the potential at the disc inner edge is given by:
\begin{equation}
\begin{aligned}
\psi(\Delta)  = -\psiout \Delta^{1+s} & \left[ \sum_{\substack{n=0\\n \ne n_\Delta}}^\infty{I_n \left(1-\Delta^{2n-s-1}\right)} \right. \\
& \qquad \qquad \qquad \left. - \gamma_{n_\Delta} \ln \Delta \right].
\end{aligned}
\label{eq:psiin2singular}
\end{equation}

In a similar way, if $s$ is such that:
\begin{equation}
\label{eq:n1}
2n+s+2=0
\end{equation}
at a rank $n \equiv n_1$, then one term in Eq. (\ref{eq:psiout2}) must be treated separately. This happens for $s \in {\cal E}_1$ where:
\begin{equation}
{\cal E}_1 = \{-2,-4,-6, -8, \dots\}.
\end{equation}
From Eq. (\ref{fn:div}), we have:
\begin{equation}
\psi(1)  = -\psiout \left[ \sum_{\substack{n=0\\n \ne n_1}}^\infty{J_n \left(1-\Delta^{2n+s+2}\right) - \gamma_{n_1} \ln \Delta} \right].
\label{eq:psiout2singular}
\end{equation}

We note that $s$ {\it cannot belong simultaneously to set ${\cal E}_\Delta$ and to set ${\cal E}_1$}. 

\subsection{Cases with $\Delta=0$}

\begin{figure}
\includegraphics[width=9.0cm]{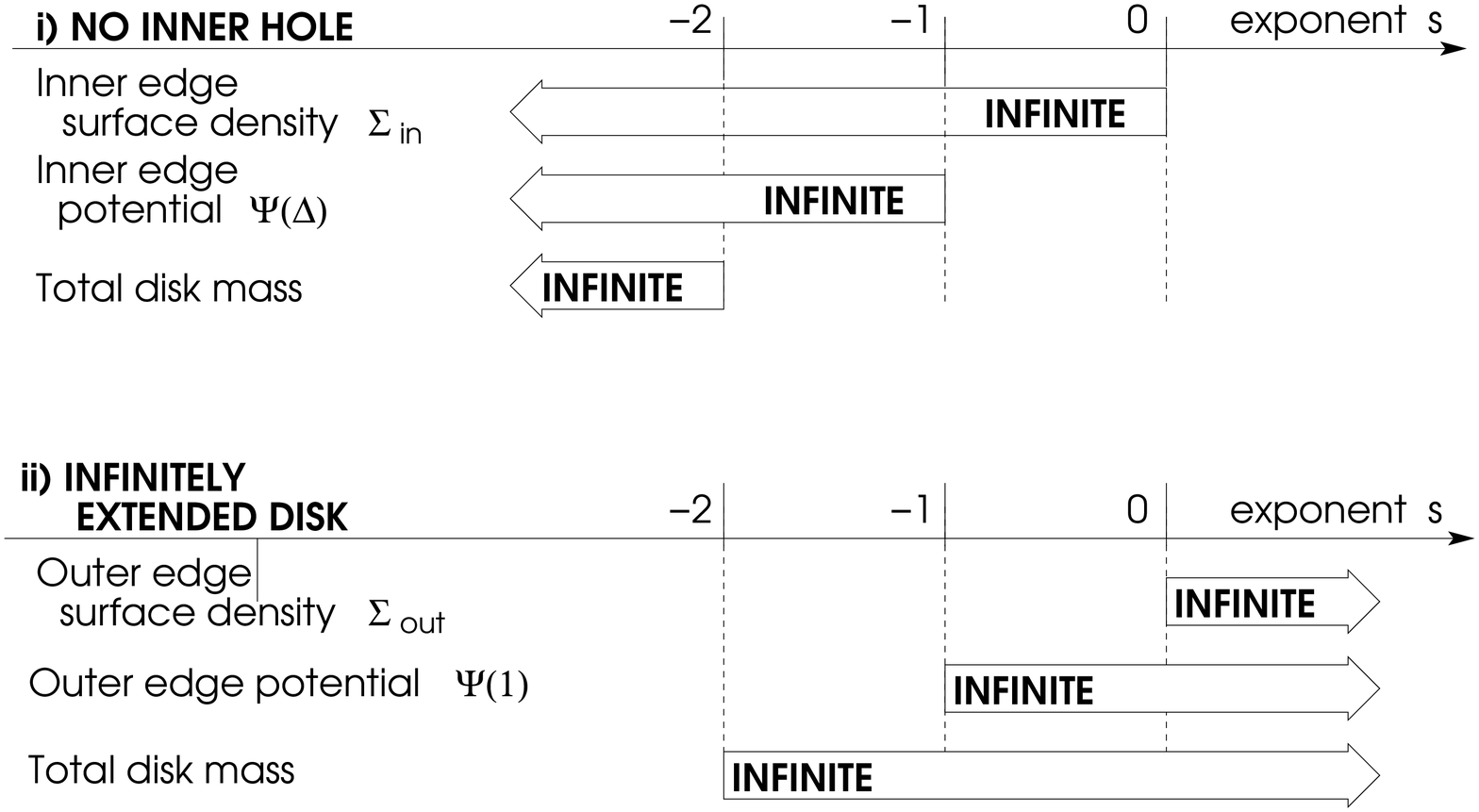}
\caption{Ranges of $s$ where the edge surface density, edge potential and total disc mass are infinite.}
\label{fig:scheme2.eps}
\end{figure}

The case $\Delta=0$ occurs i) when the disc has no inner hole (i.e. $\ain=0$) but finite size, and/or ii) when the disc has an inner edge but is infinitely extended (i.e. $\aout \rightarrow \infty$). In the first case, $\psi(\Delta)$ (denoted $\psi_c$ in Paper I) becomes the potential at the origin of coordinates  and has infinite value as soon as $1+s\le 0$. In the second case, $\psi(1)$ represents the gravitational potential at infinity. It diverges if $s+1 \ge 0$. Figure \ref{fig:scheme2.eps} summarises the ranges of $s$ where the edge surface density, edge potential, and total disc mass are finite. We note that only discs having either i) $\ain = 0$, $\aout \ne \infty$ with $s > 0$, or ii) $\ain >0$, $\aout \rightarrow \infty$ with $s \le -2$  are physically meaningful since they are characterised by a finite surface density, a finite total mass and a finite potential. Mestel discs do not belong to these categories \cite[e.g.][]{mestel63, hunter84}.

\section{Solution of the ODE in the form of series}
\label{sec:Solution of the ODE}

We see from Eqs. (\ref{eq:spiecewise}) and (\ref{eq:kseries}) that the function $S$ can  be easily expressed as a series. We have:
\begin{equation}
\label{eq:spiecewise2}
S(\varpi) = \frac{\psiout}{\varpi^2} \sum_{n=0}^\infty{\gamma_n \times }
\begin{cases}
 \varpi^{2n+1}\left( 1 - \Delta^{1+s-2n}  \right) \\  \qquad \text{in region I},\\\\
 \varpi^{2n+1} - \varpi^{-2n}\Delta^{s+2+2n} \\ \qquad \text{in region II},\\\\
  \varpi^{-2n}\left( 1 - \Delta^{s+2+2n} \right) \\ \qquad \text{in region III.}
\end{cases}
\end{equation}

By inserting this general expression into Eq. (\ref{eq:formals}), we find for $s \notin {\cal E}_\Delta \cup {\cal E}_1$ and $\Delta \ne 0$:
\begin{equation}
\begin{aligned}
\psi(\varpi) = & \frac{\psi(\varpi_0)}{\varpi_0^{1+s}} \varpi^{1+s} \\
 & +  \psiout   \sum_{n=0}^\infty{} \left[ a_n  \left( \varpi^{2n} - \frac{\varpi^{1+s}}{\varpi_0^{1+s-2n}}\right) \right.\\
 & \qquad  \qquad  \qquad  \left. + b_n \left( \frac{1}{\varpi^{2n+1}} - \frac{\varpi^{1+s}}{\varpi_0^{2n+2+s}} \right)  \right],
\end{aligned}
\label{eq:explicits}
\end{equation}
where the coefficients $a_n$ and $b_n$ are respectively given by:
\begin{equation}
\label{eq:coefapiecewise}
a_n = I_n \times
\begin{cases}
1-\Delta^{1+s-2n} \quad  \text{in region I},\\ \\
1 \quad \text{in region II},\\ \\
0 \quad \text{in region III},
\end{cases}
\end{equation}
and
\begin{equation}
\label{eq:coefbpiecewise}
b_n = J_n \times
\begin{cases}
0 \quad \text{in region I},\\ \\
\Delta^{2n+s+2} \quad \text{in region II},\\ \\
\Delta^{2n+s+2}-1 \quad \text{in region III}.
\end{cases}
\end{equation}
Since $\psi(\Delta)$ and $\psi(1)$ are available (see Sect. \ref{sec:pate}), we use $\varpi_0 = \Delta$ and $\varpi_0 = 1$ to simplify Eq. (\ref{eq:explicits}). Using Eqs. (\ref{eq:psiin2}) and (\ref{eq:psiout2}), we thus have:
\begin{equation}
\frac{\psi(\varpi)}{\psiout} = A \varpi^{1+s} +  \sum_{n=0}^\infty{\left(a_n \varpi^{2n} + \frac{b_n}{\varpi^{2n+1}}\right)},
\label{eq:psifinal}
\end{equation}
where:
\begin{equation}
\label{eq:coefcn}
A =
\begin{cases}
0,\quad \text{in region I},\\ \\
\sum_{n=0}^\infty{c_n} \quad \text{in region II},\\\\
0,\quad \text{in region III},
\end{cases}
\end{equation}
with:
\begin{equation}
c_n = -I_n -J_n.
\end{equation}
We note that $A$ is a function of $s$.

\begin{figure}
\includegraphics[width=9.0cm]{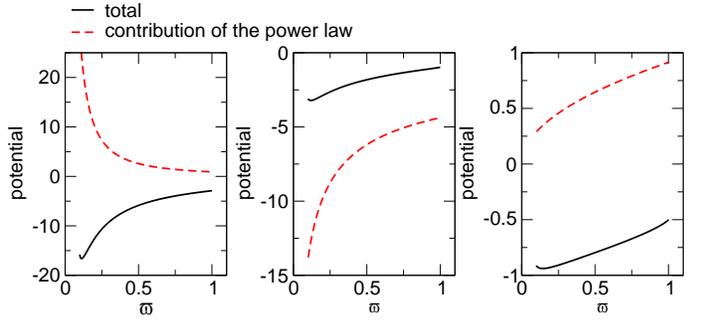}
\caption{Potential ({\it plain lines}) in a power-law disc with axis ratio $\Delta=0.1$ and $s=\{-2.5,-1.5,-0.5\}$. The contribution of the power law, i.e. the term $A \varpi^{1+s}$ in Eq. (\ref{eq:psifinal}), is shown in comparison.}
\label{fig:psin_plaw.eps}
\end{figure}

Figure \ref{fig:psin_plaw.eps} compares the total potential $\psi(\varpi)$ with the power law contribution (i.e. the term $A \varpi^{1+s}$) for three typical values of the exponent $s$ in a disc of axis ratio $\Delta = 0.1$. We clearly see that, in a finite size disc: i) the gravitational potential is not a power-law function of the radius, ii) a power-law contribution is present inside the disc only, and iii) the power law is not the dominant part of the potential. As expected, spatial self-similarity is broken due to edges.  We note that, outside the disc (i.e. in regions I and III), the series coincides with the multi-pole expansions.

\section{Potential inside the disc}
\label{sec:Potential inside the disc}

The determination of the gravitational potential is usually straightforward outside the distribution where different kinds of expansions are efficient in practice \citep{kellogg29}. In contrast, it is problematic inside matter where the classical multi-pole approach fails to converge rapidly \cite[e.g.][]{clement74,stonenorman92}.

\begin{figure}
\includegraphics[width=9.0cm]{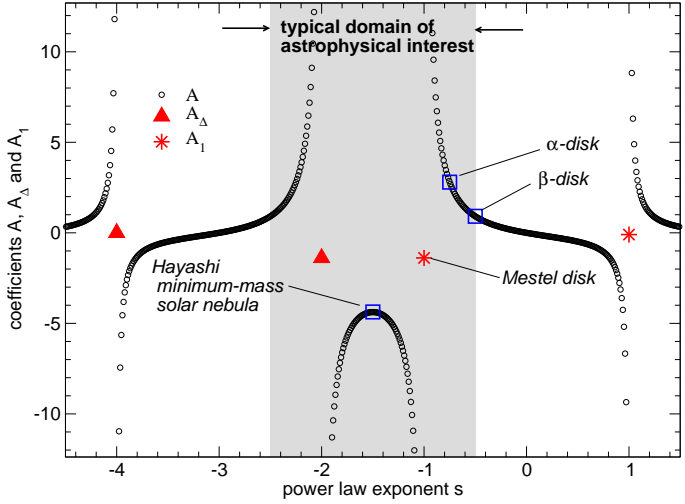}
\caption{Coefficients $A$ ({\it open circles}), $A_\Delta$  ({\it triangles}), and $A_1$ ({\it stars}) versus the power-law exponent $s$ of the surface density. A few remarkable exponents are shown ({\it open squares}): the Mestel (infinite) disc with $s=-1$ \citep{mestel63}, the Hayashi model for the solar nebula with $s=-1.5$ \citep{hayashi81}, and the $\alpha$ and $\beta$-viscosity discs where $s \approx -0.6$ depending on models \citep{ss73,cd90,dubrulle92,richardzahn99}.}
\label{fig:ainfty.eps}
\end{figure}

\subsection{Converging series}

In region II, we have $b_n = J_n \Delta^{2n+s+2}$. If we set:
\begin{equation}
X = \frac{\ain}{R} \equiv \frac{\Delta}{\varpi} \le 1,
\end{equation}
then
\begin{equation}
\frac{b_n}{\varpi^{2n+1}} = J_n X^{2n+1} \Delta^{1+s} \le  \Delta^{1+s}.
\end{equation}
 Figure \ref{fig:spectredn.eps} shows the term $a_n$ versus $n$ for $s=-1.5$ (in this case, $I_n=J_n$). As a consequence, the three series involved in Eq. (\ref{eq:psifinal}) converge rapidly since i) $I_n$ and $J_n$ both vary as $1/n^2$ at large $n$, ii) all terms are positive for large $n$, and iii) $\varpi \le 1$ and $X \le 1$. This is interesting for the truncation of series and the construction of reliable approximations (see Sect. \ref{sec:Approximate formulae}).

\subsection{The coefficient $A$}

The coefficient $A$ is plotted in Fig. \ref{fig:ainfty.eps}. It is symmetric with respect to $s=-\frac{3}{2}$ since:
\begin{equation}
\frac{c_n}{\gamma_n} = - \frac{4(4n+1)}{(4n-2s'+1)(4n+2s'+1)},
\end{equation}
where $s'=s+\frac{3}{2}$. For certain integer values of $|s|$, $A$ is strictly zero. This is in particular the case for $s=-3$ and $s=0$ (see Appendix \ref{app:azero}). As Eq. (\ref{eq:coefcn}) shows, $A$ rises as soon as the exponent $s$ is such that either $2n-s-1$ or $2n+s+2$ is small (see Sect. \ref{sec:pate}). Even, if $n = n_\Delta$ (or $n=n_1$), the coefficient $A$ apparently contains a singular term, namely $I_{n_\Delta} \varpi^{1+s}$ (resp. $J_{n_1} \varpi^{1+s}$); however, this singularity exactly cancels with the term $a_{n_\Delta} \varpi^{2 n_\Delta}$ (resp. $b_{n_1} \varpi^{-2n_1 -1}$). In practice, when $s \in {\cal E}_\Delta$, Eq. (\ref{eq:psifinal}) is no longer valid. Instead, we have:
\begin{equation}
\begin{aligned}
\frac{\psi(\varpi)}{\psiout} & = \left( A_\Delta + \gamma_{n_\Delta} \ln \varpi \right) \varpi^{1+s}\\
& \qquad +  \sum_{\substack{n=0\\n \ne n_\Delta}}^\infty{a_n \varpi^{2n}} + \sum_{n=0}^\infty{\frac{b_n}{\varpi^{2n+1}}},
\end{aligned}
\label{eq:psifinalndelta}
\end{equation}
where
\begin{equation}
A_\Delta =- \sum_{\substack{n=0\\n \ne n_\Delta}}^\infty{I_n} - \sum_{n=0}^\infty{J_n} ,
\label{eq:adelta}
\end{equation}
Similarly, if $s \in {\cal E}_1$, the potential writes
\begin{equation}
\begin{aligned}
\frac{\psi(\varpi)}{\psiout} & = \left( A_1 + \gamma_{n_1} \ln X \right) \varpi^{1+s}\\
&  \qquad +  \sum_{n=0}^\infty{a_n \varpi^{2n}}  + \sum_{\substack{n=0\\n \ne n_1}}^\infty{\frac{b_n}{\varpi^{2n+1}}},
\end{aligned}
\label{eq:psifinaln1}
\end{equation}
where
\begin{equation}
A_1=- \sum_{n=0}^\infty{I_n} - \sum_{\substack{n=0\\n \ne n_1}}^\infty{J_n},
\label{eq:a1}
\end{equation}

A few values of $A$, $A_\Delta$, and $A_1$ are listed in Appendix \ref{ax:tables}. We note that $A_\Delta$ (or $A_1$) differs only from $A$ by the term $I_{n_\Delta}$ (resp. $J_{n_1}$).

\subsection{Convergence acceleration}
\label{subsec:Convergence acceleration}

Once $s$ is given, the coefficients $A$, $A_\Delta$, and $A_1$ can be easily determined at the required accuracy. It is also possible to improve the convergence rate of the associated series. This accelerates the computation of the coefficients and makes their dependence with the exponent $s$ more explicit. Convergence acceleration is performed by using the properties of the definite integrals of the complete elliptic integrals of the first and second kinds. The demonstration reported in the Appendix \ref{ax:accconv} yields, for $s \notin {\cal E}_\Delta \cup {\cal E}_1$:
\begin{equation}
A= \frac{1-6C}{\pi} - s(s+2)\sum_{n=0}^\infty{d_n} - (s+1)(s+3)\sum_{n=0}^\infty{e_n},
\label{eq:aacc}
\end{equation}
where
\begin{equation}
\left\{
\begin{aligned}
d_n &= \frac{I_n}{4n^2-1},\\ \\
e_n &=  \frac{J_n}{4n^2-1}.
\end{aligned}
\right.
\end{equation}
and $C$ is the Catalan constant (half the area under the function $\elik$). Numerically, the constant in $A$ is
\begin{equation}
\frac{1-6C}{\pi} \approx -1.4310555380011220.
\end{equation}

Figure \ref{fig:spectredn.eps} shows the coefficient $d_n$ for $s=-1.5$ (in this case, $d_n=e_n$). It follows that, for large $n$, $d_n$ and $e_n$ behave like $\sim 1/n^4$ asymptotically, and so $A$ approaches its converged value far more rapidly than by means of Eq. (\ref{eq:coefcn}).

For $s \in {\cal E}_\Delta$, we then find:
\begin{equation}
\begin{aligned}
A_\Delta  =  \frac{1-6C}{\pi} & + \gamma_{n_\Delta} \frac{4n_\Delta}{4n_\Delta^2-1} - s(s+2) \sum_{\substack{n=0\\n \ne n_\Delta}}^\infty{d_n}\\
& -(s+1)(s+3) \sum_{n=0}^\infty{e_n},
\end{aligned}
\label{eq:aaccndelta}
\end{equation}
instead of Eq. (\ref{eq:adelta}), and for $s \in {\cal E}_1$, this is:
\begin{equation}
\begin{aligned}
A_1 = \frac{1-6C}{\pi} & + \gamma_{n_1} \frac{4n_1}{4n_1^2-1} - s(s+2)\sum_{n=0}^\infty{d_n} \\
& - (s+1)(s+3) \sum_{\substack{n=0\\n \ne n_1}}^\infty{e_n},
\end{aligned}
\label{eq:aaccn1}
\end{equation}
instead of Eq. (\ref{eq:a1}).

\begin{figure}
\includegraphics[width=8.5cm]{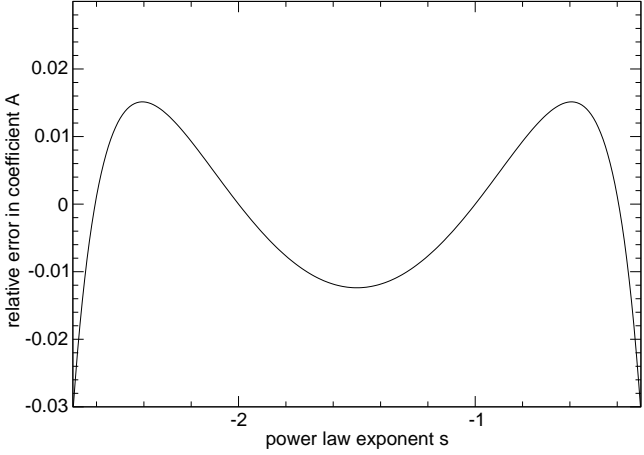}
\caption{Relative error in $A$ when computed approximately from Eq. (\ref{eq:ainftyapprox}).}
\label{fig:errainfty.eps}
\end{figure}                                                                    

Depending on the exponent $s$ of interest, a good approximation for $A$, $A_\Delta$ or $A_1$ can be obtained by considering only the largest terms in the sum, i.e. all terms up to the rank $n \approx \frac{1}{2}(1+s)$ or $n \approx -\frac{1}{2}(s+2)$. For astrophysical discs, $s$ is around $-1$ meaning that we retain only the first term in Eq. (\ref{eq:aacc}). We then find the following approximation:
\begin{equation}
A \approx \frac{1-6C}{\pi} - \frac{s(s+2)}{s+1} +\frac{(s+1)(s+3)}{s+2},
\label{eq:ainftyapprox}
\end{equation}
whose accuracy is better than $3 \%$ for $-2.7 \lesssim s \lesssim -0.3$ as Fig. \ref{fig:errainfty.eps} shows. For $s \in \{-2,-1\}$, we find from Eqs. (\ref{eq:aaccndelta}) and (\ref{eq:aaccn1}):
\begin{equation}
A_\Delta = A_1 \approx \frac{1-6C}{\pi}
\label{eq:adelta1approx}
\end{equation}
which is in good agreement with the converged value given in Table \ref{ax:tables}.

\begin{figure}
\includegraphics[width=9.0cm]{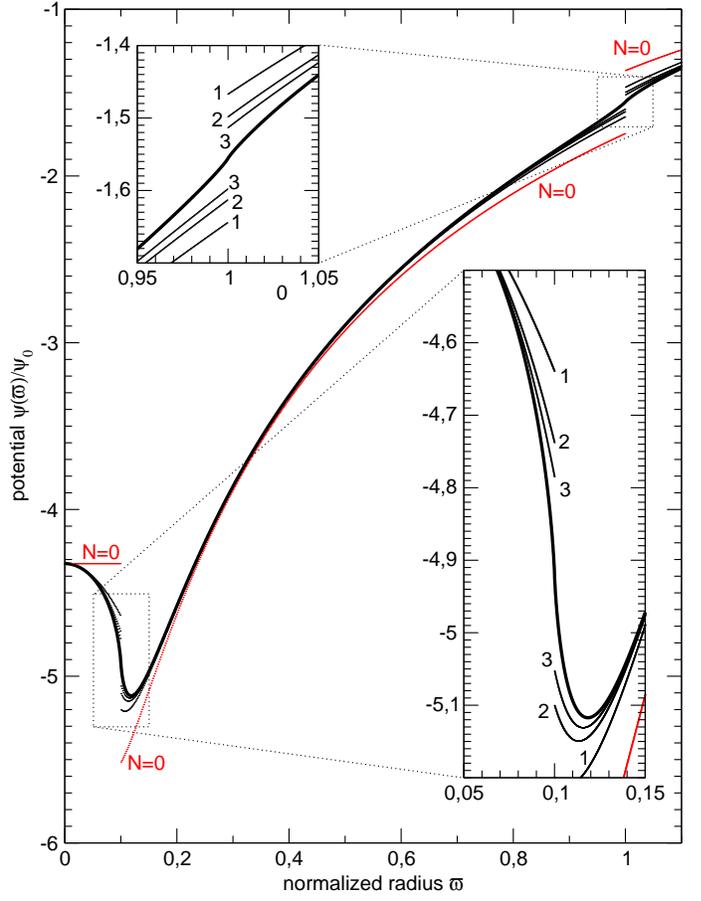}
\caption{The exact potential for a power law disc with $s=-1.5$ ({\it thick line}) compared to approximate values  ({\it thin lines}) found from Eq. (\ref{eq:psiapprox}) for $N=0$, $1$, $2$ and $3$ (i.e. see Eqs. (\ref{eq:psiapprox0}), (\ref{eq:psiapprox1}), (\ref{eq:psiapprox2}), etc.). The largest errors are found around edges.}
\label{fig:psin.eps}
\end{figure}                                                                    

\section{Approximate formulae inside the disc}
\label{sec:Approximate formulae}

Equations (\ref{eq:psifinal}), (\ref{eq:psifinalndelta}) and (\ref{eq:psifinaln1}) contain three rapidly converging series that can be truncated to derive reliable approximations for the potential. For $s \approx -1$, only a few terms can be considered (see Sect. \ref{sec:Potential inside the disc}). Although many truncations are possible, we have noticed that the most accurate approximations of $\psi$ for discs\footnote{Here, discs are supposed to be objects of axis ratio $\Delta \lesssim 0.1$.} are obtained provided the coefficient $A$ (or $A_\Delta$ or $A_1$ depending on $s$) takes its converged value. Under these circumstances, the $N$-order approximation for the potential in region II becomes:
\begin{equation}
\frac{\psiapp^{(N)}(\varpi)}{\psiout} = A \varpi^{1+s} +  \sum_{n=0}^N{\left(a_n \varpi^{2n} + \frac{b_n}{\varpi^{2n+1}}\right)},
\label{eq:psiapprox}
\end{equation}
which is, in its asymptotic limit:
\begin{equation}
\lim_{N \rightarrow \infty}\psiapp^{(N)}(\varpi) \equiv  \psi(\varpi).
\end{equation}

\subsection{Zero-order approximation}

As argued in Sect. \ref{subsec:Convergence acceleration}, a reliable formula for the potential in astrophysical discs (for which $s \approx -1.5 \pm 1$) is obtained by considering only the terms $a_0$ and $b_0/\varpi$, in addition to the power law. At the lowest order, we thus have\footnote{A careful treatment of the singular cases $s=\{-2,-1\}$ shows that Eq. (\ref{eq:psiapprox0}c) yields Eqs. (\ref{eq:psiapprox0}a) and (\ref{eq:psiapprox0}b). \label{notesing}}:
\begin{equation}
\frac{\psiapp^{(0)}(\varpi)}{\psiout} = \left\{
\begin{aligned}
&\frac{A_1 + \ln X}{\varpi} + 1 \quad \text{if}\; s = -2, \\ \\
&A_\Delta + \ln \varpi + X \quad \text{if}\; s = -1, \\ \\
&A \varpi^{1+s} -  \frac{1}{s+1}\\
& \qquad + \frac{\Delta^{1+s} X}{(2+s)} \quad \text{otherwise}.
\label{eq:psiapprox0}
\end{aligned}
\right.
\end{equation}
where, in this case, $A_1 = A_\Delta \approx -1.386$ (see Appendix \ref{ax:tables}). Figure \ref{fig:psin.eps} compares this zero-order approximation with the exact potential for typical disc parameters. It follows that the relative deviation $|1 - \psiapp^{(0)}/\psi|$ does not exceed $10 \%$, the deviation being the largest close to the edges.

\begin{figure}
\includegraphics[width=9.0cm]{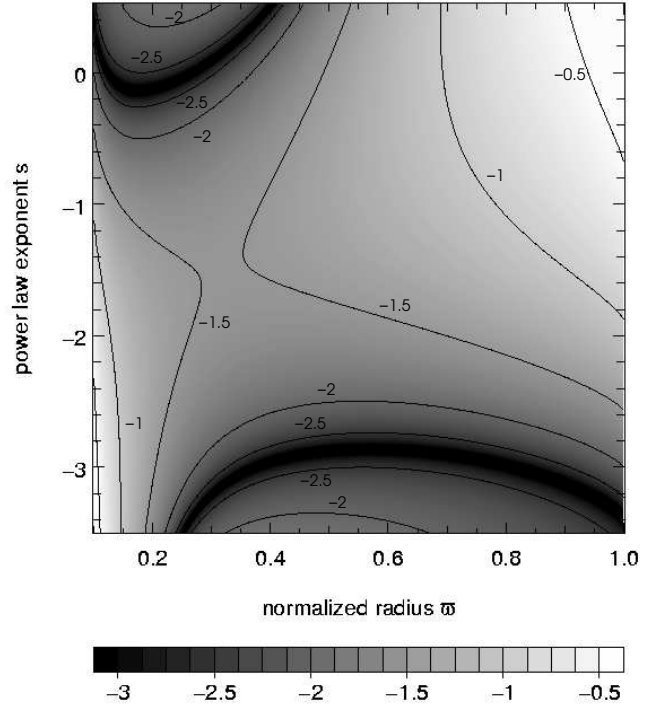}
\caption{Contour map showing the decimal logarithm of the relative error in the potential in the $(\varpi,s)-$plane when $\psi$ is approximated by Eq. (\ref{eq:psiaxwithax}).}
\label{fig:errorapprox.eps}
\end{figure}

It is worth noting that the accuracy remains of the same order if coefficient $A$ is determined by Eq. (\ref{eq:ainftyapprox}). This is convenient when the explicit dependence of $\psi$ on $s$ is required. Under this hypothesis, the potential becomes (see note \ref{notesing}):
\begin{equation}
\begin{aligned}
\label{eq:psiaxwithax}
\psi(R) & \approx - 2 \pi G \Sigout \aout \times \left\{ \frac{1}{1+s}  \right.\\
& \left. + \left[ 0.431 - \frac{1}{(1+s)(2+s)} \right] \left(\frac{R}{\aout}\right)^{1+s}  \right.\\
& \qquad \qquad \qquad \left. - \frac{1}{s+2}\left(\frac{\ain}{\aout}\right)^{1+s} \frac{\ain}{R}\right\}.
\end{aligned}
\end{equation}
Figure \ref{fig:errorapprox.eps} shows the accuracy of this formula in the $(\varpi,s)-$plane. We see that the relative deviation of $\sim 10 \%$ observed previously for $s=-1.5$ holds globally for $s$ roughly in the range\footnote{This range of exponents should be appropriate for most astrophysical applications (see the introduction).} $[-3,0]$. This agrees with the fact that Eq. (\ref{eq:ainftyapprox}) produces values of $A$ within a few percents for this range of exponents. The deviation can be reduced at the inner and outer edges provided additional terms are included (see below).

\subsection{Higher orders}

If necessary, more accurate expressions are obtained by accounting gradually for following terms (each acting as a smaller and smaller correction). For $N=1$, we have:
\begin{equation}
\frac{\psiapp^{(1)}(\varpi)}{\psiout} = \frac{\psiapp^{(0)}(\varpi)}{\psiout}  + \frac{\varpi^2}{4(1-s)}
 + \frac{\Delta^{1+s} X^3}{4(4+s)},
\label{eq:psiapprox1}
\end{equation}
and for $N=2$, this is:
\begin{equation}
\frac{\psiapp^{(2)}(\varpi)}{\psiout} = \frac{\psiapp^{(1)}(\varpi)}{\psiout}   + \frac{9 \varpi^4}{64(3-s)} + \frac{9 \Delta^{1+s} X^5}{64(6+s)},
\label{eq:psiapprox2}
\end{equation}
and so on. We note that Eq. (\ref{eq:psifinal}) is particularly well suited to numerical computation since $\psi$ can be determined by means of a recurrence procedure. A possible algorithm (not including the treatment of singular cases where $s \in {\cal E}_\Delta \cup {\cal E}_1$) is proposed in the Appendix \ref{ax:algo}.

\section{Discs with no inner/outer edges}
\label{sec:Discs with no inner/outer edges}

\subsection{Finite disc without inner hole}

If the disc has no inner edge but a finite size (i.e. $\ain=0$ and $\aout \ne \infty$), then the ODE is \cite[see][]{hh07}:
\begin{equation}
\label{eq:ssinf1}
S(\varpi) = \frac{ 2 \psiout}{\pi \varpi^2}
\begin{cases}
 \varpi \elik(\varpi), \quad \text{in region II},\\\\
 \elik\left(\frac{1}{\varpi}\right), \quad \text{in region III}.
\end{cases}
\end{equation}
It can be verified that the solution is still described by Eqs. (\ref{eq:psifinal}) and (\ref{eq:aacc}), but the coefficients $a_n$ and $b_n$ are:
\begin{equation}
a_n =
\begin{cases}
I_n, \quad \text{in region II}\\
0, \quad \text{in region III}
\end{cases}
\end{equation}
and
\begin{equation}
b_n =
\begin{cases}
0, \quad \text{in region II}\\
-J_n, \quad \text{in region III}
\end{cases}
\end{equation}

\subsection{Infinitely extended disc with inner hole}

If the disc is infinite but has an inner edge (i.e. $\ain > 0$ and $\aout \rightarrow \infty$), then
\begin{equation}
\frac{d \psi}{dY} - (1+s)\frac{\psi}{Y} = S(Y),
\end{equation}
where 
\begin{equation}
Y = \frac{R}{\ain} \equiv \frac{\varpi}{\Delta},
\end{equation}
is the new space variable,
\begin{equation}
\label{eq:ssinf2}
S(Y) = - \frac{ 2 \psiin}{\pi Y^2}
\begin{cases}
Y \elik (Y), \quad \text{in region I}\\\\
\elik\left(\frac{1}{Y}\right), \quad \text{in region II},
\end{cases}
\end{equation}
and $\psiin = 2 \pi G \Sigin \ain$ is a constant (this is not the potential at the inner edge). The analogue of Eq. (\ref{eq:psifinal}) is:
\begin{equation}
\frac{\psi(Y)}{\psiin} = A Y^{1+s} +  \sum_{n=0}^\infty{\left(a_n Y^{2n} + \frac{b_n}{Y^{2n+1}}\right)},
\label{eq:psifinaly}
\end{equation}
where $A$ is still given by Eq. (\ref{eq:coefcn}),
\begin{equation}
a_n =
\begin{cases}
- I_n, \quad \text{in region I}\\
0, \quad \text{in region II}
\end{cases}
\end{equation}
and
\begin{equation}
b_n =
\begin{cases}
0, \quad \text{in region I}\\
J_n, \quad \text{in region II}.
\end{cases}
\end{equation}

\subsection{Infinitely extended disc}

If the disc is infinite, the ODE become homogeneous:
\begin{equation}
\frac{d \psi}{dR} - (1+s)\frac{\psi}{R} = 0,
\end{equation}
and the solution is a power law:
\begin{equation}
\psi(R) = \psi(R_0) \left(\frac{R}{R_0}\right)^{1+s},
\end{equation}
where $R_0$ is some reference radius. In this case only, a self-similar surface density can rigorously be associated with a self-similar potential. The presence of edges destroys this property. We note that, for $s=-1$ (i.e. Mestel's disc), the derivation of the ODE requires a careful treatment. The integral form, i.e. Eq. (\ref{eq:psiintegralform}), gives:
\begin{equation}
\begin{aligned}
\psi(R) & = - 4 G \Sigma_0 a_0 \left\{ 2C + \frac{\pi}{2} \left[ \sum_{n=1}^\infty{\frac{\gamma_n}{2n} \left(\frac{R}{a}\right)^{2n}} \right. \right.\\
        & \qquad \qquad \qquad \qquad  \qquad \qquad \left. \left. +  \ln \frac{a}{R} \right]_{a=\aout}^{a=R} \right\}.
\end{aligned}
\end{equation}
We then have
\begin{equation}
\frac{d \psi}{d R} = 2 \pi G \Sigma_0 a_0 \frac{1}{R}.
\end{equation}
This expression is compatible with Eq. (\ref{eq:ssinf1}a) when $\aout \rightarrow \infty$ (in this case, region III no longer exists and we have $S=\psiout/\varpi$).

\section{Concluding remarks}

In this paper, we have derived an exact expression for the gravitational potential in the plane of flat power-law discs as a solution of the ODE reported in Paper I. This expression is valid over the entire spatial domain and takes into account finite size effects. Inside the disc (the most difficult case to treat in general), it consists of three terms of comparable magnitude: a power law of the cylindrical radius $R$ with index $1+s$ (where $s$ is the exponent of the surface density) and two series of $R$ and $1/R$. In terms of convergence, our expression is by far superior to the multi-pole expansion method. Reliable approximations for the potential can be produced by performing fully controlled truncations. We have shown that the potential can be expressed by means of a simple function of $R$ and $s$, which is valid to within a few percents in the range of exponents $-3 \lesssim s \lesssim 0$. This formula should be sufficiently accurate for most astrophysical applications. If necessary, more accurate formulae can be developped by including successive terms. These results should help in investigating various phenomena where disc gravity plays a significant role.

An interesting point concerns the case of discs for which the surface density is not a power-law function. As shown in Paper I, it is easy to reproduce numerically the potential when the profile $\Sigma(R)$ is a mixture of power laws. From an analytical point of view however, the construction of a reliable formula for $\psi$ as compact as the one obtained here is not guaranteed at all. For instance, for an expansion of the form:
\begin{equation}
\Sigma(R)=\sum_{s=0}^M{\alpha_s R^s},
\end{equation}
where $s$ is an integer, each of the $M+1$ series $\{a_n,b_n\}$  should be truncated at a rank $N \gtrsim n_\Delta \sim (s+1)/2$ (see Eqs. (\ref{eq:psifinal}) and (\ref{eq:psiapprox})), which corresponds to an approximate formula for $\psi$ containing about $2M(M+1)$ terms. The number of terms to consider can become prohibitively large when several power laws are required.

\begin{acknowledgements}
F. Hersant was supported by a CNRS fellowship which is gratefully acknowledged. We thank C. Baruteau. We thank the anonymous referee for valuable comments.
\end{acknowledgements}

\bibliographystyle{aa}

\begin{thebibliography}{26}
\expandafter\ifx\csname natexlab\endcsname\relax\def\natexlab#1{#1}\fi

\bibitem[{{Baruteau} \& {Masset}(2008)}]{baruteaumasset08}
{Baruteau}, C. \& {Masset}, F. 2008, ArXiv e-prints, 801

\bibitem[{{Bisnovatyi-Kogan}(1975)}]{bkogan75}
{Bisnovatyi-Kogan}, G.~S. 1975, Soviet Astronomy Letters, 1, 177

\bibitem[{{Clement}(1974)}]{clement74}
{Clement}, M.~J. 1974, \apj, 194, 709

\bibitem[{{Collin-Souffrin} \& {Dumont}(1990)}]{cd90}
{Collin-Souffrin}, S. \& {Dumont}, A.~M. 1990, \aap, 229, 292

\bibitem[{{Dubrulle}(1992)}]{dubrulle92}
{Dubrulle}, B. 1992, \aap, 266, 592

\bibitem[{{Durand}(1953)}]{durand64}
{Durand}, E. 1953, {Electrostatique. Vol. I. Les distributions.} (Ed. Masson)

\bibitem[{{Edgar}(2007)}]{edgar07}
{Edgar}, R.~G. 2007, \apj, 663, 1325

\bibitem[{{Evans}(1994)}]{evans94}
{Evans}, N.~W. 1994, \mnras, 267, 333

\bibitem[{{Evans} \& {Read}(1998)}]{evansread98}
{Evans}, N.~W. \& {Read}, J.~C.~A. 1998, \mnras, 300, 83

\bibitem[{{Gradshteyn} \& {Ryzhik}(1965)}]{gradryz65}
{Gradshteyn}, I.~S. \& {Ryzhik}, I.~M. 1965, {Table of integrals, series and
  products} (New York: Academic Press, 1965, 4th ed., edited by Geronimus,
  Yu.V.~(4th ed.); Tseytlin, M.Yu.~(4th ed.))

\bibitem[{{Hayashi}(1981)}]{hayashi81}
{Hayashi}, C. 1981, Progress of Theoretical Physics Supplement, 70, 35

\bibitem[{{Hughes} {et~al.}(2008){Hughes}, {Wilner}, {Qi}, \&
  {Hogerheijde}}]{hughes08}
{Hughes}, A.~M., {Wilner}, D.~J., {Qi}, C., \& {Hogerheijde}, M.~R. 2008, ArXiv
  e-prints, 801

\bibitem[{{Hunter} {et~al.}(1984){Hunter}, {Ball}, \& {Gottesman}}]{hunter84}
{Hunter}, Jr., J.~H., {Ball}, R., \& {Gottesman}, S.~T. 1984, \mnras, 208, 1

\bibitem[{{Hur\'e}(1998)}]{hure98}
{Hur\'e}, J.-M. 1998, \aap, 337, 625

\bibitem[{{Hur{\'e}} \& {Hersant}(2007)}]{hh07}
{Hur{\'e}}, J.-M. \& {Hersant}, F. 2007, \aap, 467, 907

\bibitem[{{Hur{\'e}} {et~al.}(2001){Hur{\'e}}, {Richard}, \& {Zahn}}]{hrz01}
{Hur{\'e}}, J.-M., {Richard}, D., \& {Zahn}, J.-P. 2001, \aap, 367, 1087

\bibitem[{{Kellogg}(1929)}]{kellogg29}
{Kellogg}, O.~D. 1929, {Foundations of Potential Theory} (New-York: Frederick
  Ungar Publishing Company)

\bibitem[{{Mestel}(1963)}]{mestel63}
{Mestel}, L. 1963, \mnras, 126, 553

\bibitem[{{Pi{\'e}tu} {et~al.}(2007){Pi{\'e}tu}, {Dutrey}, \&
  {Guilloteau}}]{pietu07}
{Pi{\'e}tu}, V., {Dutrey}, A., \& {Guilloteau}, S. 2007, \aap, 467, 163

\bibitem[{{Pringle}(1981)}]{pringle81}
{Pringle}, J.~E. 1981, \araa, 19, 137

\bibitem[{{Richard} \& {Zahn}(1999)}]{richardzahn99}
{Richard}, D. \& {Zahn}, J.-P. 1999, \aap, 347, 734

\bibitem[{{Rybicki} \& {Lightman}(1979)}]{rybicki79}
{Rybicki}, G.~B. \& {Lightman}, A.~P. 1979, {Radiative processes in
  astrophysics} (New York, Wiley-Interscience, 1979.~393 p.)

\bibitem[{{Semer{\'a}k}(2004)}]{semerak04}
{Semer{\'a}k}, O. 2004, Classical and Quantum Gravity, 21, 2203

\bibitem[{{Shakura} \& {Sunyaev}(1973)}]{ss73}
{Shakura}, N.~I. \& {Sunyaev}, R.~A. 1973, \aap, 24, 337

\bibitem[{{Stone} \& {Norman}(1992)}]{stonenorman92}
{Stone}, J.~M. \& {Norman}, M.~L. 1992, \apjs, 80, 753

\bibitem[{{Zhao} {et~al.}(1999){Zhao}, {Carollo}, \& {de Zeeuw}}]{zhao99}
{Zhao}, H., {Carollo}, C.~M., \& {de Zeeuw}, P.~T. 1999, \mnras, 304, 457

\end{thebibliography}

\onecolumn

\newpage
\appendix 

\section{Vanishing coefficient $A$ for $s\in\{-3,0\}$}
\label{app:azero}

For $s\in \{-3,0\}$, the coefficient is :
\begin{equation}
A = - \sum_0^\infty{\gamma_n \left( \frac{1}{2n+2}+\frac{1}{2n-1} \right)}.
\end{equation}
We notice that the first sum is the definite integral of $u K(u)$, which is
\begin{equation}
\begin{aligned}
\sum_0^\infty{\frac{\gamma_n}{2n+2}} & = \sum_0^\infty{\gamma_n \int_0^1{u^{2n+1}du}},\\
& = \int_0^1{ du \sum_0^\infty{\gamma_n u^{2n+1}}},\\
& = \int_0^1{u du \sum_0^\infty{\gamma_n u^{2n}du}},\\
& = \frac{2}{\pi}\int_0^1{\elik(u) u du},\\
& = \frac{2}{\pi} \left[ \elie(u) - \left(1 - u^2 \right) \elik(u) \right]_0^1,\\
& = \frac{2}{\pi}.
\end{aligned}
\end{equation}
To find the second sum, we compute the definite integral of $K(u)/u^2$ in two ways. First, we have by direct integration  \citep[e.g.][]{gradryz65}:
\begin{equation}
\begin{aligned}
\frac{2}{\pi}\int_0^1{\frac{\elik(u)}{u^2}du} & = \frac{2}{\pi} \left[ - \frac{\elie(u)}{u}\right]_0^1,\\
& = - \frac{2}{\pi} \left[ \elie(1) - \lim_{u \rightarrow 0} \frac{\elie(u)}{u} \right],\\
& = - \frac{2}{\pi} +  \frac{2}{\pi} \lim_{u \rightarrow 0} \frac{\elie(u)}{u}, \\
& = - \frac{2}{\pi} + \lim_{u \rightarrow 0} \frac{1}{u}.
\end{aligned}
\end{equation}
Second, by replacing above $\elik$ by its series representation, we also find:
\begin{equation}
\begin{aligned}
\frac{2}{\pi}\int_0^1{\frac{\elik(u)}{u^2}du} & = \int_0^1{du \sum_0^\infty{\gamma_n u^{2n-2}}},\\
& = \int_0^1{du \left( \frac{\gamma_0}{u^2} + \sum_1^\infty{\gamma_n u^{2n-2}}\right)},\\
& = - \gamma_0 \left[1 - \lim_{u \rightarrow 0} \frac{1}{u} \right] + \sum_1^\infty{\frac{\gamma_n}{2n-1}},\\
& = - \gamma_0 + \gamma_0 \lim_{u \rightarrow 0} \frac{1}{u} + \left(\frac{\gamma_0}{-1} + \sum_1^\infty{\frac{\gamma_n}{2n-1}} - \frac{\gamma_0}{-1} \right),\\
& = \gamma_0 \lim_{u \rightarrow 0} \frac{1}{u} + \sum_0^\infty{\frac{\gamma_n}{2n-1}}.
\end{aligned}
\end{equation}
Since $\gamma_0=1$, we have
\begin{equation}
 \sum_0^\infty{\frac{\gamma_n}{2n-1}} = - \frac{2}{\pi},
\end{equation}
and so $A=0$ for $s=-3$ and $s = 0$.

\newpage

\section{Converged values of coefficient $A$, $A_1$, and $A_\Delta$ for $s \in [-4,1]$}
\label{ax:tables}

\begin{table}[h]
\centering
\begin{tabular}{ccccccl}
$s$ & $-3-s$ & $s'$ & coefficient $A$ & relative error & iterations & comment \\ \hline

 {\tt -4.00} & {\tt +1.00} & {\tt -2.50} & {\tt -9.65735902799760E-02} & {\tt +3.1E-15 }& {\tt  6200} &  values of $A_\Delta$ and $A_1$\\
 {\tt -3.95} & {\tt +0.95} & {\tt -2.45} & {\tt -5.08795971306598E+00} & {\tt +1.6E-16 }& {\tt  4763}\\
 {\tt -3.90} & {\tt +0.90} & {\tt -2.40} & {\tt -2.57897148688964E+00} & {\tt +9.9E-17 }& {\tt  6289}\\
 {\tt -3.85} & {\tt +0.85} & {\tt -2.35} & {\tt -1.73622548292480E+00} & {\tt +3.9E-17 }& {\tt  8653}\\
 {\tt -3.80} & {\tt +0.80} & {\tt -2.30} & {\tt -1.30966680644896E+00} & {\tt +2.1E-17 }& {\tt 10760}\\
 {\tt -3.75} & {\tt +0.75} & {\tt -2.25} & {\tt -1.04923490407066E+00} & {\tt +8.1E-17 }& {\tt  7977}\\
 {\tt -3.70} & {\tt +0.70} & {\tt -2.20} & {\tt -8.71529232712215E-01} & {\tt +1.4E-16 }& {\tt  7154}\\
 {\tt -3.65} & {\tt +0.65} & {\tt -2.15} & {\tt -7.40760471984439E-01} & {\tt +2.1E-16 }& {\tt  6696}\\
 {\tt -3.60} & {\tt +0.60} & {\tt -2.10} & {\tt -6.38987239513759E-01} & {\tt +2.8E-16 }& {\tt  6379}\\
 {\tt -3.55} & {\tt +0.55} & {\tt -2.05} & {\tt -5.56193814181879E-01} & {\tt +3.5E-16 }& {\tt  6133}\\
 {\tt -3.50} & {\tt +0.50} & {\tt -2.00} & {\tt -4.86319914494768E-01} & {\tt +4.3E-16 }& {\tt  5926}\\
 {\tt -3.45} & {\tt +0.45} & {\tt -1.95} & {\tt -4.25454571886386E-01} & {\tt +5.2E-16 }& {\tt  5743}\\
 {\tt -3.40} & {\tt +0.40} & {\tt -1.90} & {\tt -3.70931368062394E-01} & {\tt +6.3E-16 }& {\tt  5576}\\
 {\tt -3.35} & {\tt +0.35} & {\tt -1.85} & {\tt -3.20839275213352E-01} & {\tt +7.7E-16 }& {\tt  5420}\\
 {\tt -3.30} & {\tt +0.30} & {\tt -1.80} & {\tt -2.73740796314985E-01} & {\tt +9.4E-16 }& {\tt  5270}\\
 {\tt -3.25} & {\tt +0.25} & {\tt -1.75} & {\tt -2.28500045775376E-01} & {\tt +1.2E-15 }& {\tt  5124}\\
 {\tt -3.20} & {\tt +0.20} & {\tt -1.70} & {\tt -1.84171894656878E-01} & {\tt +1.5E-15 }& {\tt  4980}\\
 {\tt -3.15} & {\tt +0.15} & {\tt -1.65} & {\tt -1.39926048201652E-01} & {\tt +2.0E-15 }& {\tt  4837}\\
 {\tt -3.10} & {\tt +0.10} & {\tt -1.60} & {\tt -9.49911954664950E-02} & {\tt +3.1E-15 }& {\tt  4694}\\
 {\tt -3.05} & {\tt +0.05} & {\tt -1.55} & {\tt -4.86101640289562E-02} & {\tt +6.3E-15 }& {\tt  4550}\\
 {\tt -3.00} & {\tt +0.00} & {\tt -1.50} & {\tt -8.70956628336228E-15} & {\tt +3.6E-02 }& {\tt  4403} & strictly $0$ (see Appendix \ref{app:azero})\\
 {\tt -2.95} & {\tt -0.05} & {\tt -1.45} & {\tt +5.16876888183838E-02} & {\tt +6.4E-15 }& {\tt  4252}\\
 {\tt -2.90} & {\tt -0.10} & {\tt -1.40} & {\tt +1.07410381098617E-01} & {\tt +3.2E-15 }& {\tt  4098}\\
 {\tt -2.85} & {\tt -0.15} & {\tt -1.35} & {\tt +1.68286941322651E-01} & {\tt +2.1E-15 }& {\tt  3937}\\
 {\tt -2.80} & {\tt -0.20} & {\tt -1.30} & {\tt +2.35664501680007E-01} & {\tt +1.6E-15 }& {\tt  3771}\\
 {\tt -2.75} & {\tt -0.25} & {\tt -1.25} & {\tt +3.11208301380443E-01} & {\tt +1.2E-15 }& {\tt  3596}\\
 {\tt -2.70} & {\tt -0.30} & {\tt -1.20} & {\tt +3.97027142932255E-01} & {\tt +1.0E-15 }& {\tt  3410}\\
 {\tt -2.65} & {\tt -0.35} & {\tt -1.15} & {\tt +4.95854560428368E-01} & {\tt +8.6E-16 }& {\tt  3212}\\
 {\tt -2.60} & {\tt -0.40} & {\tt -1.10} & {\tt +6.11318936974807E-01} & {\tt +7.4E-16 }& {\tt  2997}\\
 {\tt -2.55} & {\tt -0.45} & {\tt -1.05} & {\tt +7.48359798181055E-01} & {\tt +6.5E-16 }& {\tt  2759}\\
 {\tt -2.50} & {\tt -0.50} & {\tt -1.00} & {\tt +9.13893162088934E-01} & {\tt +5.7E-16 }& {\tt  2486}\\
 {\tt -2.45} & {\tt -0.55} & {\tt -0.95} & {\tt +1.11792019234452E+00} & {\tt +5.1E-16 }& {\tt  2151}\\
 {\tt -2.40} & {\tt -0.60} & {\tt -0.90} & {\tt +1.37546760819327E+00} & {\tt +4.5E-16 }& {\tt  1662}\\
 {\tt -2.35} & {\tt -0.65} & {\tt -0.85} & {\tt +1.71019225943660E+00} & {\tt +4.1E-16 }& {\tt  1335}\\
 {\tt -2.30} & {\tt -0.70} & {\tt -0.80} & {\tt +2.16159222263118E+00} & {\tt +3.7E-16 }& {\tt  1821}\\
 {\tt -2.25} & {\tt -0.75} & {\tt -0.75} & {\tt +2.80087471242395E+00} & {\tt +3.4E-16 }& {\tt  1997}\\
 {\tt -2.20} & {\tt -0.80} & {\tt -0.70} & {\tt +3.77063202688011E+00} & {\tt +3.1E-16 }& {\tt  2058}\\
 {\tt -2.15} & {\tt -0.85} & {\tt -0.65} & {\tt +5.40388067136072E+00} & {\tt +2.8E-16 }& {\tt  2036}\\
 {\tt -2.10} & {\tt -0.90} & {\tt -0.60} & {\tt +8.70024086898869E+00} & {\tt +2.6E-16 }& {\tt  1928}\\
 {\tt -2.05} & {\tt -0.95} & {\tt -0.55} & {\tt +1.86592556634413E+01} & {\tt +2.4E-16 }& {\tt  1681}\\
  {\tt -2.00} & {\tt -1.00} & {\tt -0.50} & {\tt -1.38629436111989E+00} & {\tt +7.2E-18 }& {\tt  7955} &  values of $A_\Delta$ and $A_1$ \\
 {\tt -1.95} & {\tt -1.05} & {\tt -0.45} & {\tt -2.14370823367682E+01} & {\tt +2.1E-16 }& {\tt  1770}\\
 {\tt -1.90} & {\tt -1.10} & {\tt -0.40} & {\tt -1.14939346514465E+01} & {\tt +1.9E-16 }& {\tt  2141}\\
 {\tt -1.85} & {\tt -1.15} & {\tt -0.35} & {\tt -8.22454216651923E+00} & {\tt +1.8E-16 }& {\tt  2399}\\
 {\tt -1.80} & {\tt -1.20} & {\tt -0.30} & {\tt -6.63018820764787E+00} & {\tt +1.7E-16 }& {\tt  2598}\\
 {\tt -1.75} & {\tt -1.25} & {\tt -0.25} & {\tt -5.71250114438470E+00} & {\tt +1.7E-16 }& {\tt  2755}\\
 {\tt -1.70} & {\tt -1.30} & {\tt -0.20} & {\tt -5.14024384191494E+00} & {\tt +1.6E-16 }& {\tt  2878}\\
 {\tt -1.65} & {\tt -1.35} & {\tt -0.15} & {\tt -4.77330268633758E+00} & {\tt +1.6E-16 }& {\tt  2972}\\
 {\tt -1.60} & {\tt -1.40} & {\tt -0.10} & {\tt -4.54390925876444E+00} & {\tt +1.5E-16 }& {\tt  3039}\\
 {\tt -1.55} & {\tt -1.45} & {\tt -0.05} & {\tt -4.41737401180816E+00} & {\tt +1.5E-16 }& {\tt  3078}\\
 {\tt -1.50} & {\tt -1.50} & {\tt +0.00} & {\tt -4.37687923045294E+00} & {\tt +1.5E-16 }& {\tt  3091}\\
 {\tt -1.45} & {\tt -1.55} & {\tt +0.05} & {\tt -4.41737401180816E+00} & {\tt +1.5E-16 }& {\tt  3078}\\
 \hline
\end{tabular}
\caption{Values of $A$ ($4$th column) computed within the computer precision (double precision; about $16$-digit) for different power law exponents $s$ or $s'$ (columns $1$ to $3$) including those relevant for astrophysical applications (see also Fig. \ref{fig:ainfty.eps}). Also given are the relative error on the coefficient ($5$-th column) and the number of iterations ($6$-th column) required.}
\label{table:ainfty}
\end{table}

\newpage
\section{Convergence acceleration for $A$}
\label{ax:accconv}

The coefficient $A$ is given by a series whose convergence can be accelerated by considering an interesting property of the integral of $\elik$ \citep[e.g.][]{gradryz65}, namely :
\begin{equation}
\int_0^1{\elik(x)dx} = \frac{\pi}{2}\sum_0^\infty{\frac{\gamma_n}{2n+1}} = 2C,
\label{eq:integk}
\end{equation}
where $C$ is Catalan's constant. We can then write
\begin{equation}
\begin{aligned}
\sum_{n=0}^\infty{\frac{\gamma_n}{1+s-2n}}  & = \sum_{n=0}^\infty{\gamma_n\left(\frac{1}{1+s-2n}+\frac{1}{2n+1}\right)} - \sum_{n=0}^\infty{\frac{\gamma_n}{2n+1}}\\
 & = (s+2)  \sum_{n=0}^\infty{\frac{\gamma_n}{(2n+1)(1+s-2n)}} - \frac{4C}{\pi},
\end{aligned}
\end{equation}
where terms vary like $1/n^3$ for large $n$. A second convergence acceleration can be obtained by considering the complete elliptic integral of the second kind : 
\begin{equation}
\elie(x)=\int_0^{\pi/2}{d \phi \sqrt{1-x^2 \sin^2 \phi}}, \quad 0 \le x \le 1,
\end{equation}
whose definite integral over the modulus $x$ is :
\begin{equation}
\int_0^1{\elie(x)dx} = -\frac{\pi}{2}\sum_0^\infty{\frac{\gamma_n}{(2n+1)(2n-1)}} = \frac{1}{2}+C.
\label{eq:intege}
\end{equation}
We then have :
\begin{equation}
\begin{aligned}
\sum_{n=0}^\infty{\frac{\gamma_n}{(2n+1)(1+s-2n)}} & =  \sum_{n=0}^\infty{\gamma_n\left[\frac{1}{(2n+1)(1+s-2n)} +\frac{1}{(2n+1)(2n-1)}\right]} - \sum_{n=0}^\infty{\frac{\gamma_n}{(2n+1)(2n-1)}}\\
& = s \sum_{n=0}^\infty{\frac{\gamma_n}{(2n+1)(1+s-2n)(2n-1)}} +\frac{2}{\pi}\left(\frac{1}{2}+C \right).
\end{aligned}
\end{equation}
It follows that :
\begin{equation}
\begin{aligned}
\sum_{n=0}^\infty{\frac{\gamma_n}{1+s-2n}}  & = s(s+2) \sum_{n=0}^\infty{\frac{\gamma_n}{(2n+1)(1+s-2n)(2n-1)}}  + \frac{2+s+2sC}{\pi}.
\end{aligned}
\end{equation}

The second term in Eq. (\ref{eq:coefcn}) gives :
\begin{equation}
\begin{aligned}
\sum_{n=0}^\infty{\frac{\gamma_n}{2n+s+2}} & = \sum_{n=0}^\infty{\gamma_n\left(\frac{1}{2n+s+2}-\frac{1}{2n+1}\right) + \sum_{n=0}^\infty{\frac{\gamma_n}{2n+1}}}\\
 & = -(s+1)  \sum_{n=0}^\infty{\frac{\gamma_n}{(2n+1)(2n+s+2)}} + \frac{4C}{\pi}.
\end{aligned}
\end{equation}

But, from Eq. (\ref{eq:intege}), we have :
\begin{equation}
\begin{aligned}
 \sum_{n=0}^\infty{\frac{\gamma_n}{(2n+1)(2n+s+2)}} & =  \sum_{n=0}^\infty{\gamma_n\left[\frac{1}{(2n+1)(2n+s+2)}-\frac{1}{(2n+1)(2n-1)}\right]} + \sum_{n=0}^\infty{\frac{\gamma_n}{(2n+1)(2n-1)}}\\
 & = -(s+3)\sum_{n=0}^\infty{\frac{\gamma_n}{(2n+1)(2n-1)(2n+s+2)}} - \frac{2}{\pi}\left(\frac{1}{2}+C \right),
\end{aligned}
\end{equation}
and then :
\begin{equation}
\begin{aligned}
\sum_{n=0}^\infty{\frac{\gamma_n}{2n+s+2}} & = (s+1)(s+3)  \sum_{n=0}^\infty{\frac{\gamma_n}{(2n+1)(2n-1)(2n+s+2)}} + \frac{1+s+2(s+3)C}{\pi}
\end{aligned}
\end{equation}
Finally, we have :
\begin{equation}
\begin{aligned}
A & =  \sum_{n=0}^\infty{c_n}\\
  & =  s(s+2) \sum_{n=0}^\infty{\frac{\gamma_n}{(2n+1)(1+s-2n)(2n-1)}}  -(s+1)(s+3)  \sum_{n=0}^\infty{\frac{\gamma_n}{(2n+1)(2n-1)(2n+s+2)}} + \frac{1-6C}{\pi}
\end{aligned}
\end{equation}

\section{A basic algorithm}
\label{ax:algo}

A possible algorithm is the following (see the text for cases where  $s \in {\cal E}_1 \cup {\cal E}_\Delta$):

\begin{enumerate}
\item \underline{disc parameters} (see Sect. \ref{sec:unified}):\\ \\
- set the inner edge $\ain=???$\\
- set the outer edge $\aout=???$\\
- set the outer surface density $\Sigout=???$\\
- set the power law exponent $s=???$\\ \\
\quad $\rightarrow$ compute $\Delta$\\
\item \underline{power-law coefficient $A$} :\\ \\
\quad $\rightarrow$ compute $A$ (or $A_\Delta$ or $A_1$ depending on $s$) at the {\it computer accuracy} (see Sect. \ref{sec:Approximate formulae}); see Eqs. (\ref{eq:aacc}), (\ref{eq:aaccndelta}) and (\ref{eq:aaccn1}), or use {\it pre-computed converged values} (see Appendix \ref{ax:tables})\\
\item \underline{radius}\\\\
- set the radius $R = ???$\\ \\
\quad $\rightarrow$ compute dimensionless variables $\varpi$ and $X$ (depending on the region I, II or III)\\
\item \underline{initializations}\\ \\
- set $\gamma_0$ to $1$; see Eqs. (\ref{eq:kseries})\\
- set $a_0$ to $\frac{-1}{1+s}$; see Eq. (\ref{eq:coefapiecewise})\\
- set $b_0$ to $\frac{1}{2+s}$; see Eq. (\ref{eq:coefbpiecewise})\\
- set $\psi$ to $0$\\
\item \underline{main loop} on $n$\\ \\
\quad $\rightarrow$ compute $\gamma_n$ from $\gamma_{n-1}$\\
\quad $\rightarrow$ compute $a_n$ from $a_{n-1}$\\
\quad $\rightarrow$ compute $b_n$ from $b_{n-1}$\\
\quad $\rightarrow$ compute $\varpi^{2n}$ and $X^{2n+1}$\\
\quad $\rightarrow$ update $\psi$ to $\psi+ a_n \varpi^{2n} + b_n \varpi^{-2n-1}$; see Eq. (\ref{eq:psiapprox})\\
\item \underline{power-law contribution}\\ \\
\quad $\rightarrow$ update $\psi$ to $\psi+ A  \varpi^{1+s}$; see Eq. (\ref{eq:psiapprox})\\
\item \underline{final potential value}\\ \\
\quad $\rightarrow$ change $\psi$ for $\psi \times 2 \pi G \Sigout \aout$
\end{enumerate}

The loop ends after $N$ steps; the accuracy of the potential value is then given by the next term (rank $N+1$).

\end{document}